\documentclass{article}
\pdfoutput=1 
\usepackage[english]{babel}
\usepackage[utf8x]{inputenc}
\usepackage[T1]{fontenc}
\usepackage[a4paper,top=3cm,bottom=2cm,left=3cm,right=3cm,marginparwidth=1.75cm]{geometry}
\usepackage{amsmath}
\usepackage{graphicx}

\usepackage{hyperref}
\usepackage[version=3]{mhchem}
\usepackage{siunitx}

\title{Broadband photoacoustic spectroscopy of \ce{^14CH4} with a high-power mid-infrared optical frequency comb}
\date{}
\author{}

\begin{document}
\maketitle
\begin{centering}
\vspace{-40pt}
Juho Karhu\textsuperscript{1}, Teemu Tomberg\textsuperscript{1}, Francisco Senna Vieira\textsuperscript{1,2}, Guillaume Genoud\textsuperscript{3}, Vesa Hänninen\textsuperscript{1}, Markku Vainio\textsuperscript{1,4}, Markus Metsälä\textsuperscript{1}, Tuomas Hieta\textsuperscript{5}, Steven Bell\textsuperscript{6}, and Lauri Halonen\textsuperscript{1,*}\\
\vspace{15pt}
\textsuperscript{1}Department of Chemistry, University of Helsinki, P.O. Box 55, FI-00014, Helsinki, Finland\\
\textsuperscript{2}Currently with the Department of Physics, Umeå University, 901 87 Umeå, Sweden\\
\textsuperscript{3}Centre for Metrology MIKES, VTT Technical Research Centre of Finland Ltd, P.O. Box 1000, VTT, 02044, Finland\\
\textsuperscript{4}Photonics Laboratory, Physics Unit, Tampere University, FI-33014 Tampere, Finland\\
\textsuperscript{5}Gasera Ltd., Lemminkäisenkatu 59, FI-20520,
Turku, Finland\\
\textsuperscript{6}National Physical Laboratory, Hampton Road, Teddington, Middlesex, TW11 0LW, UK\\
\textsuperscript{*}Corresponding author: lauri.halonen@helsinki.fi\\
\end{centering}
\vspace{5pt}
\hrule
\vspace{-2pt}
\begin{abstract}
\noindent We report a photoacoustic spectroscopy setup with a high-power mid-infrared frequency comb as the light source. The setup is used in broadband spectroscopy of radiocarbon methane. Due to the high sensitivity of a cantilever-enhanced photoacoustic cell and the high power light source, we can reach a detection limit below 100 ppb in a broadband measurement with a sample volume of only a few milliliters. The first infrared spectrum of \ce{^14CH4} is reported and given a preliminary assignment. The results lay a foundation for the development of optical detection systems for radiocarbon methane.\\
\\
\noindent©2019 Optical Society of America. One print or electronic copy may be made for personal use only. Systematic reproduction and distribution, duplication of any material in this paper for a fee or for commercial purposes, or modifications of the content of this paper are prohibited.\\
\url{https://doi.org/10.1364/OL.44.001142}
\end{abstract}
\hrule
\vspace{6pt}
Cantilever-enhanced photoacoustic spectroscopy (CEPAS) is a versatile technique for spectroscopic sensing in gas and condensed phase. In the gas phase, CEPAS has been used in sensitive trace gas measurements, with detection limits down to ppt level and below \cite{Peltola,Tomberg}. The most sensitive methods use single mode lasers, but CEPAS has also been used as a detector for broadband Fourier transform infrared (FTIR) measurements \cite{Uotila}. Recently, broadband CEPAS setups, which use modern coherent broadband sources, namely supercontinua and optical frequency combs (OFC), have been demonstrated \cite{Mikkonen,Sadiek}. The main advantages of CEPAS detection are that the sensitivity scales linearly with the power of the light source, which can potentially lead to a large dynamic range, and that the sample volume can be kept small while still achieving high sensitivity. In this letter, we use a high-power singly resonant femtosecond optical parametric oscillator (fs-OPO) as the light source in FTIR-CEPAS to reach the best sensitivity of a broadband photoacoustic measurement to date. We use the setup in a measurement of the first reported infrared spectrum of \ce{^14CH4}.

The best known application for measuring the concentration of the radioactive \ce{^14C} isotope is carbon dating. Determining the \ce{^14C}/\ce{^12C}-ratio can also be used to differentiate between biogenic and fossil emission, since in the latter the \ce{^14C} concentration is depleted \cite{Mohn}. The current standard for sensitive detection of \ce{^14C} is accelerator mass spectrometry (AMS), but it requires large and costly facilities to operate. Lately, there has been vast interest in the development of optical detection of radiocarbon compounds \cite{Galli,Genoud,Fleisher,McCartt,Sonnenschein}. Optical detectors are generally more affordable and compact alternatives, allowing the possibility of field measurements. A sensitive laboratory setup, where the detection limit is approaching the sensitivity of AMS, has been demonstrated in the case of \ce{^14CO2} \cite{Galli}. 

The success with the optical \ce{^14CO2} detectors shows potential for the development of optical detectors for other radiocarbon compounds. Radiocarbon methane is a clear candidate for spectroscopic detection, as it is an important small molecule. Methane is the main component of natural gas and a potent greenhouse gas. Monitoring of the \ce{^14CH4} content can be used to determine the biofraction in biogas samples or to apportion \ce{CH4} sources. The radiocarbon isotopologue \ce{^14CH4} is known to be emitted by light water nuclear reactors \cite{Yim}. As an organic compound, radioactive \ce{CH4} is more harmful than \ce{^14CO2}, since it is more easily absorbed by living organisms, and its monitoring is important in the context of nuclear facilities. The development of optical detectors for \ce{^14CH4} is hindered by the lack of spectral information. While the infrared spectra of the stable \ce{CH4} isotopologues are well studied, the spectrum of \ce{^14CH4} is largely unknown. In this letter, we use FTIR-CEPAS for a broadband measurement of the antisymmetric CH-stretching vibrational band $\nu_3$ of \ce{^14CH4}. The \ce{^14CH4} lines are indentified according to known spectral features of the other \ce{CH4} isotopes and a theoretical calculation of the isotope shift. The \ce{^14C} content in available samples is often restricted, mainly due to radiation safety standards. Photoacoustic spectroscopy is well suited for measurements with limited sample volume, since the photoacoustic signal is inversely proportional to the sample volume \cite{Kreuzer}. The high sensitivity allows for the use of a sample with low activity concentration. 

\begin{figure}[tbp]
\centering
\includegraphics[width=0.8\linewidth]{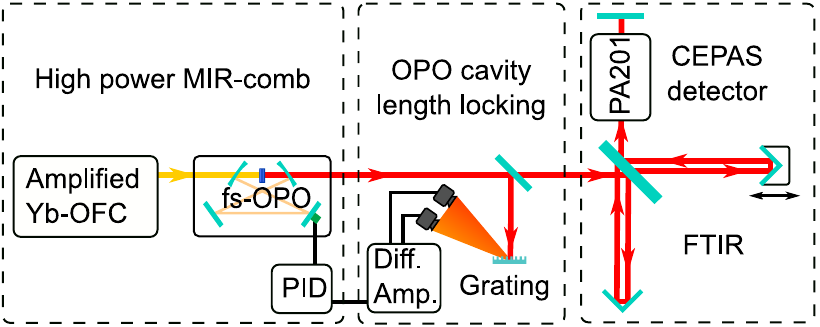}
\caption{Schematic picture of the measurement setup. An amplified ytterbium frequency comb pumps an fs-OPO. The fs-OPO idler is used as a light source for FTIR-CEPAS. A PID controller tunes one of the fs-OPO cavity mirror positions with a piezo actuator to fix the difference between the intensities at two wavelengths.}
\label{fig:scheme}
\end{figure}

Figure \ref{fig:scheme} shows the experimental setup. The light source is the idler beam from an fs-OPO. The fs-OPO is pumped with an amplified ytterbium OFC, which has a center wavelength at about 1040 nm (Orange Comb, Menlo Systems) and maximum output power of 10 W. The nonlinear medium is a 10 mm long periodically poled lithium niobate crystal, doped with magnesium oxide. The fs-OPO is singly resonant, with the signal beam resonating inside the cavity. The signal and idler wavelengths are about 1.5 and \SI{3}{\micro\m}, respectively. The fs-OPO cavity length is stabilized with a simple feedback circuit to ensure a constant spectral envelope for the idler \cite{Wachman}. Part of the output idler beam is sampled with a pellicle beam splitter and dispersed with a grating. Two parts of the dispersed beam are focused to two different photodetectors. The difference between the intensities measured at the two wavelengths is stabilized by tuning one of the fs-OPO cavity mirrors with a piezo actuator. The carrier-envelope offset frequency of the fs-OPO frequency comb remains unlocked. The fs-OPO idler beam is directed to a commercial high-resolution FTIR spectrometer (FS120, Bruker). A commercial CEPAS cell (PA201, Gasera) is used as a detector. The interferograms are measured symmetrically and the power spectrum is calculated to avoid phase errors. The maximum resolution of a two-sided measurement is \SI{0.02}{\per\cm}. The bandwidth of the CEPAS detector is limited to about \SI{700}{Hz}. Therefore, a slow scanning speed of 0.14 cm/s, in terms of the optical path difference, was used with the FTIR. A single two-sided scan with the highest resolution takes about 12 min to measure.

The gas sample was a mixture of concentrated \ce{^14CH4} (Quotient Bioresearch) and nitrogen. The total \ce{CH4} concentration was about 100 ppm, with about 1 ppm of \ce{^14CH4} isotopologue. The gas lines and the CEPAS cell were flushed with nitrogen gas prior to measurements. The gas cell was also flushed briefly with the sample gas, before a static sample was placed into the PA201 cell for the measurement. The sample volume inside the cell is about \SI{8}{\milli\l}. The temperature of the cell was fixed to \SI{25}{\celsius}.

In addition to the photoacoustic spectrum, the spectrum of the fs-OPO is recorded with the same spectrometer, using an MCT detector (PVI-2TE-5, VIGO). The photoacoustic signal is proportional to the spectral power density, so the raw photoacoustic spectrum was divided by the fs-OPO spectrum to cancel out the power dependence. Other effects to the spectral response, for example any frequency dependence of the photoacoustic sensitivity, were ignored, since they should be small when the measurement is done outside the resonance frequency of the cantilever microphone. A low resolution (\SI{1}{\per\cm}) spectrum was used for the normalization, similarly to the procedure described in \cite{Sadiek}. In this way, no additional noise was introduced to the photoacoustic spectrum by the normalization. However, this may cause local discrepancies in the intensities, because some sharp features of the fs-OPO spectrum, such as water absorption from the laboratory air, are not fully resolved.

\begin{figure}[tbp]
\centering
\includegraphics[width=0.8\linewidth]{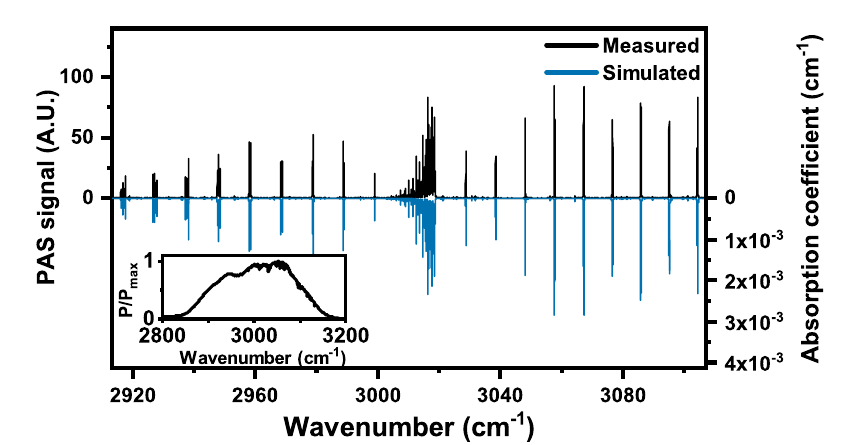}
\caption{Photoacoustic FTIR spectrum (left axis, upwards) with a simulated absorption spectrum (right axis, downwards). The measured spectrum is an average of three scans. The simulation is calculated with data from the HITRAN database. The intensity of the measured spectrum was scaled according to the normalized spectrum of the fs-OPO idler (inset). }
\label{fig:spectrum_full}
\end{figure}

Figure \ref{fig:spectrum_full} shows the measured photoacoustic spectrum. The total pressure of the sample was 250 mbar. Therefore, the pressure broadened linewidth is only about 1.5 times the instrument resolution. The spectrum is an average of three measurement scans. The power of the fs-OPO idler beam entering the CEPAS cell for these scans was about 95 mW. The measured spectrum has been scaled by the spectrum of the fs-OPO (Fig. \ref{fig:spectrum_full} inset). The main features of the measured photoacoustic spectrum are due to the stable isotopologue \ce{^12CH4}, since its concentration in the sample is about 100 times that of \ce{^14CH4}. There is also some water as impurity, mostly due to minor leaks and outgassing in the gas exchange system. Figure \ref{fig:spectrum_full} also shows, for a comparison, a simulated absorption coefficient of a gas mixture of 100 ppm of \ce{CH4} with the natural ratios of the stable isotopologues, and 570 ppm of water at a total pressure of \SI{250}{\milli\bar} (simulated with HITRAN on the Web \cite{BABIKOV}).  The \ce{^12CH4} line positions from the calculated spectrum were used to calibrate the wavelength axis of the measured spectrum.

Looking at the spectrum of the main isotopologue \ce{^12CH4}, the signal-to-noise ratio (SNR), defined as the height of the highest peak in the R-branch over one standard deviation of the background noise, was 1300 at its best for a single scan without averaging. Since the concentration of \ce{^12CH4} is about 100 ppm, the noise equivalent detection limit is about 83 ppb. The spectrum has a weak broadband background feature, the magnitude of which follows approximately the spectrum of the fs-OPO. We believe that this is mostly due to absorption by the cell windows. The noise level within this background feature is about two times the noise level without a light source. Calculating the power spectrum of a symmetric interferogram introduces nonlinearity to the noise amplitude, which also results in a small positive offset throughout the recorded spectrum.

\begin{figure}[tbp]
\centering
\includegraphics[width=0.8\linewidth]{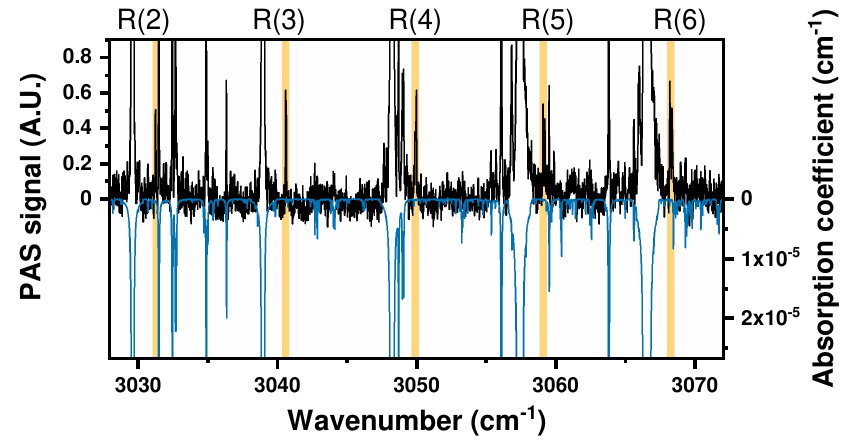}
\caption{Measured spectrum and simulation. This is a close up of Fig. \ref{fig:spectrum_full}. A constant background of 0.2 was subtracted from the measured spectrum. In the measured spectrum, a number of lines that do not appear in the calculated spectrum of water and the stable isotopologues of \ce{CH4} are marked with yellow highlights. These match well with the expected form of the R-branch of \ce{^14CH4}, following the assignment shown above the measured spectrum. }
\label{fig:spectrum_R-branch}
\end{figure}

Additional weak transitions, absent in the literature data of \ce{CH4} or water, are visible in the spectrum. Figure \ref{fig:spectrum_R-branch} is a magnification of the R-branch of the $\nu_3$ fundamental band of \ce{CH4} from Fig. \ref{fig:spectrum_full}. It shows the strongest unknown peaks present in the spectrum. The unassigned lines match well with the expected features of the \ce{^14CH4} R-branch. Both the relative intensities, and the shapes of the lines, which are split into closely packed line components due to Coriolis interactions, follow the features of the R-branch of \ce{^12CH4}, with an isotope shift of about \SI{18}{\per\cm}. For example, the unknown line at \SI{3040}{\per\cm} is similar in shape to the R(3) line of \ce{^12CH4} at about \SI{3057.7}{\per\cm}. We have calculated the vibrational term values of the $\nu_3$ band of \ce{^12CH4} and \ce{^14CH4} using the local mode model \cite{Halonen}, to estimate the expected isotope shift. The potential energy parameters were obtained from  a least squares fit to experimental vibrational data of \ce{^12CH4}. The computational values were \SI{3021.8}{\per\cm} and \SI{3000.7}{\per\cm} for \ce{^12CH4} and \ce{^14CH4}, respectively, which results in a theoretical isotope shift of \SI{21}{\per\cm}. This matches the observation within the accuracy of the model. The isotope shift is small compared to the vibrational energies, so the strength of the Coriolis coupling between the different vibrational states is affected little. Therefore, the features caused by the Coriolis splitting remain similar in the different isotopologues, as is also the case with \ce{^13CH4} \cite{McDowell}. Additionally, the substitution of the center atom affects the rotational constant only slightly. Therefore, the $\nu_3$ bands of all the different carbon isotopologues of \ce{CH4} follow similar overall trends.

\begin{figure}[tbp]
\centering
\includegraphics[width=0.8\linewidth]{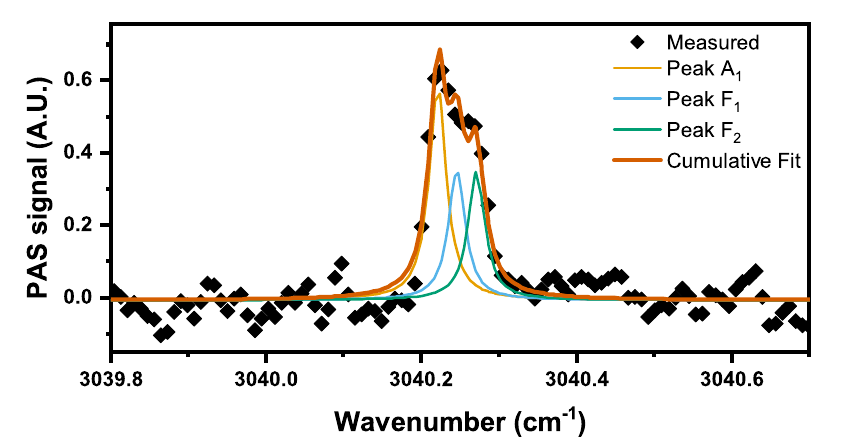}
\caption{Fit of three Lorentzian peaks on the R(3) line of \ce{^14CH4}. The relative intensities of the three lines were fixed according to the spin statistical weights. Methane R(3) line has one line component of A symmetry and two F components, and each of the two F line components have 3/5 times the area of the A component.}
\label{fig:spectrum_R3}
\end{figure}

The signal-to-noise ratio of some of the R-branch lines was high enough so that the position of the split line components could be estimated with a least-squares fit of the line shapes. The number and symmetry species of the component lines are the same as for \ce{^12CH4} and \ce{^13CH4} \cite{McDowell}. Because the SNR was relatively low and the component lines in the R-branch overlap a lot in the used pressure, a simplified model was used: The line shape of each component line was a Lorentzian-profile, since the pressure broadening is the largest contributor to the linewidth. The following constraints were also implemented: The relative intensities between the component lines were fixed to those given by the statistical weights between the symmetry species (5 for A, 3 for F, and 2 for E) \cite{Herzberg}, and the Coriolis components of the same line were constrained to have the same linewidths. To reduce the interference from nearby water and \ce{^12CH4} lines, they were included in the fitting. Water and weak \ce{^12CH4} lines were fitted with Voigt profiles, with a fixed calculated value for the Doppler width and initial fit parameter values set to HITRAN data. The strong \ce{^12CH4} lines were fitted using Hartmann-Tran profiles \cite{Ngo}. The molecular speed dependences of the linewidth and line shift, the correlation parameter, the line center position, and an intensity scaling parameter were used as fitting parameters, while the speed independent parameters of the width and shift were fixed to their HITRAN values, to speed up the calculation. Figure \ref{fig:spectrum_R3} shows a fitted profile to the strongest observed \ce{^14CH4} line. For this line, the linewidths of the fitted Lorentzian profiles were \SI{0.013}{\per\cm}. The expected pressure broadened linewidths for the same lines in \ce{^12CH4} at this pressure are about \SI{0.015}{\per\cm} according to the HITRAN database. The \ce{^14CH4} line positions from the fits are presented in the table \ref{tab:assign}. The statistical uncertainty of the fits and the low resolution limit the accuracy of the line positions.

The lines of the P-branch of \ce{^14CH4} are weaker and therefore hard to assign convincingly. Moreover, the fs-OPO power spectral density was lower on the P-branch side. In an effort to help to identify the weaker lines in the P-branch, we fitted a quadratic polynomial to the line centers that we were able to determine from the R-branch. Ignoring the Coriolis splitting and centrifugal distortion, the R and P-branches of the band obey a simple quadratic formula \cite{Herzberg}:
\begin{equation}
\bar{\nu} = \bar{\nu}_0+(B'+B''-2B'\zeta )m+(B'-B'' )m^2
\end{equation}
The symbol $\bar{\nu}$ is a position of a line, $\bar{\nu}_0$ is the vibrational band center, $B'$ and $B''$ are the effective rotational constants for the excited and ground vibrational states, respectively, and the term $2B'\zeta$ accounts for the vibrational angular momentum of the excited vibrational state. When considering the R-branch, $m$ is the rotational quantum number of the upper state.  The line positions used in the fit were calculated as the centers of gravity of the component lines. The component lines of symmetry A, E and F were given weights of 1, 2 and 3, respectively, due to their degeneracy. The results obtained with the linear least-squares method are given in table \ref{tab:bandpar}. 

\begin{table}[bp]
\centering
\caption{Band parameters in \si{\per\cm} for the $\nu_3$ band of \ce{^14CH4}}
\begin{tabular}{cc}
\hline
$\bar{\nu}_0$&	3000.621(5)\textsuperscript{a}\\
$(B'+B''-2B'\zeta )$&	10.081(2)\\
$(B'-B'' )$&	-0.0432(2)\\
\hline
\end{tabular}
\\
\textsuperscript{a}Numbers in the parenthesis are one-standard errors in the least significant digits, as given by the least-squares fit.\\
  \label{tab:bandpar}
\end{table}

In agreement with the fit, there is a clear Q-branch at about \SI{3000}{\per\cm}, but it overlaps severely with the P(2) line of \ce{^12CH4}. With the assistance of the approximate line position calculated from the fit, some lines of the P-branch can be recognized in the spectrum. Since the Coriolis components are more separated in the P-branch, a measurement at higher pressure of 1 bar, and lower resolution of \SI{0.05}{\per\cm} was performed, to obtain a slightly better SNR. However, we could assign only two lines of the P-branch. Usually the SNR was too low, or the lines overlapped with the strong \ce{^12CH4} spectrum. The P-branch lines assigned from the higher pressure spectrum are included in table \ref{tab:assign}.

In conclusion, we have demonstrated the advantage of employing a high power frequency comb in photoacoustic spectroscopy, allowing a broadband measurement with high sensitivity using a small sample volume of a few milliliters. We could reach a detection limit below 100 ppb. This is, to our knowledge, the lowest detection limit reached with FTIR photoacoustic spectroscopy up to this date, even when normalized by the measurement time. Due to a small sample volume, the OFC-CEPAS is well suited for broadband spectroscopy of samples with limited availability, such as the radioisotope sample we have used here. We have presented, to our knowledge, the first measurements of the infrared spectrum of \ce{^14CH4}. We have assigned the strongest lines of the $\nu_3$ band of \ce{^14CH4}, which do not overlap with the lines of the main isotopologue.  A pure sample without \ce{^12CH4} would lead to a more complete assignment of the \ce{^14CH4} spectrum, but such a sample was unavailable to us at this time. In any case, \ce{^12CH4} is the likely interference in most applications and our results should be of assistance in further development of optical methods for the \ce{^14CH4} detection, for example, when choosing appropriate single mode lasers for more sensitive and compact measurement setups.
\vspace{5pt}

{\bf Funding.} CHEMS doctoral school of University of Helsinki; Jenny and Antti Wihuri Foundation; the Academy of Finland (Grants 294752, 314363); European Metrology Research Programme (EMRP) project ENG54 “Metrology for Biogas”. The EMRP is jointly funded by the EMRP participating countries within EURAMET and the European Union.

\begin{table}[htb]
\centering
\caption{Preliminary assignment of the \ce{^14CH4} lines}
\begin{tabular}{cccc}
\hline
$J''$  \textsuperscript{a}& $J'$ & Symmetry\textsuperscript{b} & Line position (\si{\per\cm}) \\
\hline
4&	3&	\ce{F1}&	2959.80(4)\textsuperscript{c}\\
4&	3&	E&			2959.91(9)\\
4&	3&	\ce{F2}&	2960.00(9)\\
4&	3&	\ce{A2}&	2960.06(5)\\
3&	2&	\ce{A1}&	2970.070(14)\\
3&	2&	\ce{F1}&	2970.14(3)\\
3&	2&	\ce{F2}&	2970.25(2)\\
2&	3&	\ce{F1}&	3030.468(5)\\
2&	3&	E&			3030.485(7)\\
3&	4&	\ce{A1}&	3040.221(2)\\
3&	4&	\ce{F1}&	3040.246(3)\\
3&	4&	\ce{F2}&	3040.271(3)\\
4&	5&	\ce{F1}&	3049.901(5)\\
4&	5&	E&			3049.944(9)\\
4&	5&	\ce{F2}&	3049.974(8)\\
4&	5&	\ce{A2}&	3050.005(4)\\
5&	6&	\ce{F2}&	3059.472(4)\\
5&	6&	E&			3059.502(6)\\
5&	6&	\ce{F1}&	3059.585(4)\\
5&	6&	\ce{F2}&	3059.624(4)\\
6&	7&	\ce{A2}&	3068.967(4)\\
6&	7&	\ce{F2}&	3068.992(8)\\
6&	7&	\ce{F1}&	3069.019(6)\\
6&	7&	\ce{A1}&	3069.121(3)\\
6&	7&	\ce{F1}&	3069.151(9)\\
6&	7&	E&			3069.171(11)\\
\hline
\end{tabular}
\\
\textsuperscript{a} $J''$ and $J'$ are the rotational quantum numbers for the lower and upper state, respectively.\\
\textsuperscript{b} Symmetry label for the upper state\\
\textsuperscript{c} Numbers in the parenthesis are one-standard errors in the least significant digits, as given by the least-squares fit.\\
  \label{tab:assign}
\end{table}

\bibliographystyle{ieeetr}
\bibliography{sample}

\end{document}